\documentstyle[12pt]{article}
\newcommand{\be}{\begin{equation}}
\newcommand{\ee}{\end{equation}}
\begin{document}
\title{New DNLS Equations for Anharmonic Vibrational Impurities}
\author{{\bf M. I.  Molina$^{\dagger}$}
\vspace{1 cm}
\and
\and
Facultad de Ciencias, Departamento de F\'{\i}sica, Universidad de Chile\\
Casilla 653, Las Palmeras 3425, Santiago, Chile.\\
mmolina@abello.dic.uchile.cl}
\date{}
\maketitle
\baselineskip 18 pt
\begin{center}
{\bf Abstract}
\end{center}
\noindent
We examine some new DNLS-like equations that arise when considering strongly-coupled
electron-vibration systems, where the local oscillator potential is anharmonic. In particular,
we focus on a single, rather general nonlinear vibrational impurity and determine its bound
state(s) and its dynamical selftrapping properties.
\vspace{1cm}

\noindent
$^{\dagger}$email: mmolina@abello.dic.uchile.cl
\vspace{2 cm}

\noindent
PACS number(s):\ \ 71.55.-i

\newpage
\noindent

The standard Discrete Nonlinear Schr\"{o}dinger (DNLS) equation is a paradigmatic equation
that describes wave propagation in nonlinear and disordered systems\cite{abdullaev}, polaron
formation and dynamics in deformed media\cite{katja}, power-switching in coupled nonlinear
fibers\cite{jensen} and even tunneling dynamics of Bose-Einstein condensates inside
magneto-optical traps\cite{smerzi}. In a condensed matter context, DNLS is usually derived
from a coupled system consisting of an electron (or excitation), described quantum
mechanically, that propagates in a discrete medium while interacting strongly with
vibrational degrees of freedom, described by Einstein harmonic oscillators. In the limit where
one assumes the oscillators completely enslaved to the electrons, one arrives at the
DNLS equation:
\be
i\ \left( {d c_{n}\over{d t}} \right) = V( c_{n+1} + c_{n-1} ) - \chi | c_{n} |^{2}\ c_{n}\hspace{2cm}(\hbar \equiv 1),\label{eq:0}
\ee
where $c_{n}$ is the electronic probability amplitude at site $n$, $V$ is the
nearest-neighbor hopping parameter and $\chi$ is the nonlinearity parameter,
proportional to the square of the electron-vibration coupling\cite{jpcm_3}. The
most remarkable feature of Eq.(\ref{eq:0}) is the existence of a critical
nonlinearity parameter $\chi_{c}$, marking the onset of electronic selftrapping.
The precise value
of $\chi_{c}$ depends upon the geometry and dimensionality of the lattice, but
displays a rather universal behavior when properly scaled\cite{universal}.

Since most molecular degrees of freedom alway posses some degree of anharmonicity, it is
interesting to examine how Eq.(\ref{eq:0}) is modified when one considers weakly anharmonic
and fully anharmonic oscillators. In this work, we investigate new DNLS-like equations obtained
for three different choices of oscillator potentials: (i)\ weakly anharmonic: $U(x) = (1/2) k x^{2} +
(1/4) k_{3} x^{4}$, with $k_{3}\ll k$ (ii) Hard anharmonic: $ U(x) = k \ ( \cosh(x) - 1 )$
and (iii) Soft anharmonic:
$U(x) = (k/3) [\  u^{2} + |u| - \log(1 + |u|)\  ]$. In particular, we focus on the case of a single
vibrational impurity, and examine its bound state and dynamical selftrapping properties and compare
them to the ones of the standard DNLS case.

\section{Generalized DNLS Equations}

Let us consider an electron (or excitation) propagating in a onedimensional tight-binding
lattice, and interacting with local vibrational degrees of freedom modelled as
identical, anharmonic Einstein oscillators at each site:
\be
i \left( {d c_{n}\over{d t}} \right)  = V ( c_{n+1} + c_{n-1}) + \alpha\ u_{n}\ c_{n},\label{eq:1}
\ee
\be
m \left( {d^{2} u_{n}\over{d t^{2}}} \right) = - \left( {d U\over{d u_{n}}} \right) - \alpha\ | c_{n} |^{2},\label{eq:2}
\ee
where $\alpha$ is the electron-vibration coupling, $m$ is the oscillator mass and $U(u)$ is a
general, anharmonic potential energy. In the $m\rightarrow 0$ limit (``antiadiabatic'' limit),
the vibrations become completely enslaved to the electron and Eq.(\ref{eq:2}) becomes
\be
\left( {d U\over{d u_{n}}} \right) = - \alpha\ | c_{n} |^{2}.\label{eq:3}
\ee

After inverting Eq.(\ref{eq:3}), one obtains the oscillator equilibrium displacement in terms of
the electronic probability, 
$u_{n}(t) = u_{n}(\alpha |c_{n}|^{2})$. Inserting this back
into Eq.(\ref{eq:1}), leads to
\be
i \left( {d c_{n}\over{d t}} \right)  = V ( c_{n+1} + c_{n-1}) + \alpha\ u_{n}(\alpha | c_{n} |^{2})\ c_{n},\label{eq:4}
\ee
which constitutes formally a modified DNLS equation. Its specific form depends on
the anharmonic potential chosen. In the special case of a harmonic potential
$U(x) = (1/2) k x^{2}$, (\ref{eq:4}) reduces to the well-known standard DNLS equation (\ref{eq:0}), with
$\chi \equiv (\alpha^{2}/k)$.

\section{Single Impurity}

In the case of a single vibrational impurity, located at the origin, Eq.(\ref{eq:4}) simplifies to
\be
i \left( {d c_{n}\over{d t}} \right)  = V ( c_{n+1} + c_{n-1}) + \alpha\ u_{0}(\alpha | c_{0} |^{2})\ c_{0}\ \delta_{n,0}.
\label{eq:5}
\ee
We seek to determine the bound state(s) of Eq.(\ref{eq:5}). Taking $c_{n} = \exp( -i E t ) \ \phi_{n}$, (\ref{eq:5})
becomes
\be
(E/V) \phi_{n} = \phi_{n+1} + \phi_{n-1} + (\alpha/V)\ u_{0}(\alpha |c_{0}|^{2})\ \phi_{0}\ \delta_{n,0}.\label{eq:6}
\ee
The simplest procedure now is to assume a solution of the form $\phi_{n} = A \eta^{|n|}$,
where $A$ and $\eta$ are real and $|\eta|<1$. (Another, elegant method, is the use of lattice
Green functions\cite{green, pre}).  At the impurity location, we have:
$(E/V) = 2 \eta + (\alpha/V) u_{0}(\alpha A^{2})$, while outside the
impurity site, we have
$(E/V) = \eta + 1/\eta$. After equating these two expressions, we obtain an equation for
$\eta$:
\be
\eta^{2} + \eta\ (\alpha/V) u_{0}(\alpha A^{2}) - 1 = 0,\label{eq:7}
\ee
with solution
\be
\eta = -\left( {\alpha\over{2 V}} \right) u_{0}(\alpha A^{2}) \pm \sqrt{1 + \left( {\alpha\over{2 V}}\right)^{2} u_{0}^{2}(\alpha A^{2})}
\label{eq:8}
\ee
Without loss of generality, we will take $\alpha >0$, (i.e., $u_{0}(\alpha A^{2}) < 0$). In order
to have $|\eta| < 1$, we have to take the negative root in (\ref{eq:8}).
The unnormalized bound state profile then reads:
\be
\phi_{n} = A \left[ -\left({\alpha\over{2 V}}\right) u_{0}(\alpha A^{2}) -
\sqrt{ 1 + \left({\alpha\over{2 V}}\right)^{2} u_{0}^{2}(\alpha A^{2})}\right]^{|n|}\label{eq:9}
\ee
The other equation needed comes from the normalization condition:
$ 1 = \sum_{-\infty}^{\infty} | \phi_{n} |^{2}$. This provides the following relation between $\eta$ and $A$:
\be
\eta = -\sqrt{{1 - A^{2}\over{1 + A^{2}}}}\label{eq:10}
\ee
where $\eta$ is given by (\ref{eq:8}). After equating (\ref{eq:8}) and (\ref{eq:10}), one obtains the following
equation for $A$:
\be
-\left( {1 - A^{2}\over{1 + A^{2}}} \right)^{1/2} =
-\left( {\alpha\over{2 V}} \right) u_{0}(\alpha A^{2}) -
\left[ 1 + \left( {\alpha\over{2 V}}\right)^{2} u_{0}^{2}(\alpha A^{2}) \right]^{1/2}.\label{eq:A}
\ee
Once $A$ ($0 < A < 1$) is determined, we can know $u_{0}(\alpha A^{2})$ and then $\phi_{n}$, the bound state through
(\ref{eq:9}) and the bound state energy $E$, from $(E/V) = \eta + (1/\eta)$, or
\be
{E\over{V}} = -2 \sqrt{1 + \left({\alpha\over{2 V}}\right)^{2} u_{0}^{2}(\alpha A^{2}) },\label{eq:11}
\ee
which always lies outside the band. Since we will consider oscillator potentials that 
increase monotonically with displacement, $| u_{0}(\alpha A^{2}) |$ will always increase with
increasing coupling causing, according to (\ref{eq:11}), an increasing detachment of the bound state energy
from the band, with increasing coupling. On the other hand, we can write the bound state as
$\phi_{n} = A\ \exp(-|n|/\lambda)$, with
\be
\lambda = -\left\{ \log\left|\ -\left({\alpha\over{2 V}}\right) u_{0}(\alpha A^{2}) -
\sqrt{1 + \left( {\alpha\over{2 V}} \right)^{2} u_{0}^{2}(\alpha A^{2})}\ \right| \right\}^{-1}
\ee
is the so-called ``localization length''. Then, as the coupling is increased, $\lambda$ decreases, giving
rise to a more localized probability profile.

\section{Special Cases}

We consider now three, representative anharmonic potentials, characterized by
$U(u)$ monotonically increasing with $|u|$ and giving rise to restoring forces
$F(u)$ that are unbounded functions of $u$ from above or below. This ensures that
(\ref{eq:3}) has always a real solution for the oscillator equilibrium displacement
at the impurity site. In addition, we choose anharmonic potentials that are
completely ``soft'' or ``hard'', that is, they always lie below or above the harmonic
potential, for any displacement. With these conditions, we can predict the general
features of the selftrapping behavior: Since $U_{\mbox{soft}}<U_{\mbox{harmonic}}<U_{\mbox{hard}}$,
this means that the oscillator equilibrium displacement at the impurity site obeys:
$u_{0}^{\mbox{hard}}<u_{0}^{\mbox{harmonic}}<u_{0}^{\mbox{soft}}$. This implies that, the effective
electron-vibration term in (\ref{eq:5}) is smaller (larger) for the hard (soft) case than
for the harmonic one. In other words, the {\em effective coupling} is smaller (larger) for the
hard (soft) case than for the usual, harmonic one.

(1)\ {\em Weakly anharmonic potential}: $U(u) = (1/2) k u^{2} + (1/4) k_{3} u^{4}$.
From (\ref{eq:3}) one obtains at the impurity site: $k u_{0} + k_{3} u_{0}^{3} = -\alpha A^{2}$,
where $A = \phi_{0}$ and we assume $k_{3}/k \ll 1$.
Using a perturbative expansion, we have to first order in
$k_{3}/k$,
\be
u_{0} \approx -\left({\alpha\over{k}}\right) A^{2} \left[ 1 -
\left({k_{3}\over{k}}\right)\left({\alpha\over{k}}\right)^{2} A^{4} \right].\label{eq:u0}
\ee
A simple graphical analysis of (\ref{eq:A}), with $u_{0}$ given by (\ref{eq:u0}), reveals the
existence of a solution for $A$, $0 < A < 1$, provided that $\alpha > \sqrt{2}$ ($V = 1 = k$).
Figure 1 shows the electronic bound state probability profile $| \phi_{n} |^{2}$ for a
small positive (negative) $k_{3}/k$, corresponding
to a weakly hard (weakly soft) underlying anharmonic oscillator.
The harmonic ($k_{3} = 0$) is also shown,
for comparison. For a fixed coupling $\alpha$, the bound state is wider (narrower)
for the hard (soft) case than for the purely harmonic case. This is to be expected on general
grounds, as explained before, since for a fixed coupling $\alpha$, the effective coupling is
smaller (larger) in the hard (soft) case than in the harmonic one.

Using (\ref{eq:u0}) and the usual definition of the nonlinearity parameter $\chi \equiv \alpha^{2}/k$,
the DNLS equation corresponding to this case is given by
\be
i\ \left( {d c_{n}\over{d t}} \right) = V( c_{n+1} + c_{n-1} ) -
\chi [\ 1 - (k_{3}/k^{2}) \chi | c_{0} |^{2}\ ]\ | c_{0} |^{2}\ c_{0}\ \delta_{n,0}\label{eq:weak}.
\ee

(2) {\em Hard monotonically increasing potential}:\ $U(u) = k ( \cosh(u) - 1)$. This is an example of a
``hard'' potential that is always greater than the harmonic one, approaching the harmonic case only in the
limit of small displacement. From (\ref{eq:3}) one obtains:
$u_{0} = - \sinh((\alpha/k) A^{2})$. After inserting this into (\ref{eq:A}), and performing some
simplification, one arrives at the following nonlinear equation for $A^{2}$:
\be
A^{2} = {\alpha\over{2 V}} \sqrt{1 - A^{4}} \sinh^{-1} ((\alpha/k) A^4)\label{eq:A1}
\ee
Simple analysis of (\ref{eq:A1}) reveals the existence of a single, nonzero solution,
provided $\alpha > \sqrt{2}$ ($V = 1 = k$).  
The corresponding DNLS equation reads:
\be
i\ \left( {d c_{n}\over{d t}} \right) = V( c_{n+1} + c_{n-1} ) -
\alpha \sinh^{-1}((\alpha/k)\ |C_{0}|^2)\ c_{0}\ \delta_{n,0}\label{eq:hard}.
\ee

(3) {\em Soft monotonically increasing potential}:\ $U(u) = (k/3) [u^{2} + |u| - \log(1 + |u|) ]$. This is
an example of a ``soft'' potential that is always smaller than the harmonic one, approaching the
harmonic case only in the $u\rightarrow 0$ limit. From (\ref{eq:3}) one obtains:
$u_{0} = (3/4)[\ -(\alpha/k)A^{2} + 1 - \sqrt{1+(2/3)(\alpha/k)A^{2}+(\alpha/k)^{2}A^{4}}\ ]$.
After inserting this into (\ref{eq:A})
and after some simplification, the equation for $A$ reads ($V = 1 = k$):
\be
A^{2} = {3\over{8}} \alpha \sqrt{1 - A^{4}} \left[\ \alpha A^{2} -
1 + \sqrt{\ 1 + {2\over{3}} \alpha A^2 + \alpha^{2} A^{4}\ }\  \ \ \right]\label{eq:A2}
\ee
Now the situation is more complex, since a graphical examination of (\ref{eq:A2}) reveals the existence
of two critical coupling values $\alpha_{c}^{(1)} = 1.36964$ and $\alpha_{c}^{(2)} = \sqrt{2}$, marking
the boundary of three coupling regimes, with different number of bound states: For $\alpha<\alpha_{c}^{(1)}$,
there is no bound state, at $\alpha = \alpha_{c}^{(1)}$ there is exactly a single bound state, with a finite
amplitude. For
$\alpha_{c}^{(1)}<\alpha<\alpha_{c}^{(2)}$, there are two bound states. As $\alpha$ increases towards
$\alpha_{c}^{(2)}$, one of the solutions approaches zero, while the other remains finite. Finally, at
$\alpha = \alpha_{c}^{(2)}$ and beyond, there is a single bound state, whose localization length
decreases with increasing coupling $\alpha$. The DNLS equation for this case is
\be
i\ \left( {d c_{n}\over{d t}} \right) = V( c_{n+1} + c_{n-1} ) -
\alpha [ (\alpha/k)|c_{0}|^{2} -1 + \sqrt{1 + (\alpha/k)^{2} |C_{0}|^{4}} ]\ c_{0}\ \delta_{n,0}\label{eq:soft}.
\ee

It is interesting to point out that the conditions for the existence of bound state(s) for our
new DNLS equations (\ref{eq:hard}) and (\ref{eq:soft}), differ significatively from the ones found in
a previous investigation of bound states of a DNLS equation with a nonlinear term of the form
$\chi |C_{n}|^{\beta}\ C_{n}$. There, the ``soft'' case $0 < \beta < 2$ always displays a single bound state,
while the ``hard'' case with $\beta > 2$, has a minimum nonlinearity threshold for $\chi$ below which no
bound state exists and above which there are two bound states\cite{pre}. Now, as the reader can
easily verify, this nonlinearity term $|C_{n}|^{\beta}$ can be thought of as originating from the (impurity) potential
$U = (k/\gamma)\ | u |^{\gamma}$, with $\gamma \equiv 1+(2/\beta)$. Thus, $\beta < 2$ corresponds to
$\gamma > 2$, that is, a {\bf hard} case. Similarly, $\beta >2$ corresponds to $\gamma < 2$, a {\bf soft case}.
In this light, the old results and the current ones look rather similar.

Figure 2 shows the bound state energies for the hard, harmonic and soft cases, as a function of
the coupling $\alpha$. For the harmonic and hard cases the bound state
energy curve starts at $\alpha = \alpha_{c}^{(2)} = \sqrt{2}$ and increases monotonically in magnitude
with increasing coupling.
For the soft case, the eigenenergy curve starts earlier, at $\alpha = \alpha_{c}^{(1)} = 1.36964$.
For coupling values between this and $\alpha_{c}^{(2)}$, we have two eigenenergy curves,
one of which moves away from the band upon increasing coupling, while the other approaches
the band edge, reaching it exactly at $\alpha = \alpha_{c}^{(2)}$, as the inset in Fig. 2 shows. 
Figure 3 shows the bound state probability profile for the hard and soft cases, for a common
coupling value $\alpha > \alpha_{c}^{(2)}$. Like the weak anharmonic case,
the hard(soft) case has a larger (smaller) localization length than the usual harmonic case.


We now compare the selftrapping transition for the three DNLS-like
equations (\ref{eq:weak}), (\ref{eq:hard}) and (\ref{eq:soft}). We place the
electron initially completely on the impurity site ($n=0$) and observe its
time evolution. In particular, we focus on the long-time average probability
at the impurity site,
\be
\langle P \rangle \equiv \lim_{T\rightarrow \infty}\ \ (1/T)\ \int_{0}^{T} | C_{0}(t) |^{2}\ dt.
\ee
We use a fourth-order Runge-Kutta scheme whose precision is monitored through probability
conservation $\sum_{n} |c_{n}(t)|^{2} = 1$, and a self-expanding lattice to avoid undesired
end effects.
Figure 4 shows $\langle P \rangle$ for the weakly anharmonic case, showing three curves corresponding to
a weakly hard, harmonic and weakly soft oscillator. The standard selftrapping transition
that occurs at $\alpha^{2}/k V \approx 3.2$ for $k_{3} = 0$, is shifted to higher (lower)
values for a small  positive (negative) $k_{3}$. This is consistent with the notion of an ``effective''
nonlinearity $\chi_{\mbox{eff}} = \chi [\ 1 - (k_{3}/k^{2}) \chi |c_{0}|^{2} \ ]$ in (\ref{eq:weak}).
The selftrapping transition occurs when $\chi_{\mbox{eff}}\approx 3.2$.   Finally, in Fig. 5
we show $\langle P \rangle$ for the fully anharmonic cases (\ref{eq:hard}) and (\ref{eq:soft}), along
with the standard (harmonic) case, for comparison. Like in the weak anharmonic case, the
selftrapping transition is shifted to higher values for the hard case, while for the soft case
it is shifted to a lower threshold. We also note that in this case, the selftrapping transition
is more abrupt than in the other cases\cite{comment}.

\section{Conclusion}

We have examined some new, DNLS-like equations that arise when considering
the propagation of an electron in a onedimensional lattice, while strongly
interacting with local vibrational degrees of freedom modelled as anharmonic
Einstein oscillators. We focused on the problem of a single
vibrational impurity, and computed the bound state(s) and the dynamical
selftrapping properties. We found that, in general, the character of the
anharmonicity affects the critical electron-vibrational coupling parameter
needed to produce a bound state and to effect a selftrapping transition:\
For a hard (soft) anharmonicity, the minimum coupling needed for a bound state
to exist is larger (smaller) than for the usual, harmonic case. Similarly,
the minimum nonlinearity needed for selftrapping is larger (smaller) in the
hard (soft) case than in the harmonic one. These results can be explained on
the base of an ``effective'' electron-vibration coupling that depends only
on the magnitude of the oscillator equilibrium displacement at the impurity site,
which in turns, depends on the relative hardness of the anharmonic potential.

\vspace{2cm}

\centerline{ACKNOWLEDGMENTS}
\vspace{1cm}

\noindent
This work was supported in part by FONDECYT grant 1020139.

\newpage

\newpage

\centerline{{\bf Captions List}}
\vspace{2cm}

\noindent {\bf Fig.1 :}\ \ Weakly anharmonic vibrational impurity:\ \ Electronic bound state probability profile,
for a small hard ($k_{3} > 0$) and soft ($k_{3} < 0$) anharmonicity. The harmonic case ($k_{3} = 0$) is also shown, for
comparison. ($V = 1, k = 1, \alpha = 2$ and $k_{3}/k = 0.2$).
\vspace{0.4cm}

\noindent{\bf Fig.2 :}\ \  Fully anharmonic vibrational impurity: Electronic bound state energy
as a function of the electron-vibration coupling, for the fully hard and soft cases  considered in
section 3 ($V = 1, k = 1$). The usual harmonic case is also shown, for comparison.
\vspace{0.4cm}

\noindent{\bf Fig.3 :}\ \ Fully anharmonic vibrational impurity:\ \ Electronic bound state probability profile for
the fully hard and soft cases considered in section 3 ($V = 1, k = 1$ and $\alpha = 2$). 
\vspace{0.4cm}

\noindent{\bf Fig.4 :}\ \ Weakly anharmonic vibrational impurity: Long-term average probability for finding the
electron at the vibrational impurity site, as a function of the nonlinearity parameter $\chi = \alpha^{2}/k $. 
\vspace{0.4cm}

\noindent{\bf Fig.5:}\ \ Same as in Fig. 4, but for the fully anharmonic case.

\newpage

%
\begin{figure}[h]
\begin{center}
\leavevmode
\hbox{
\includegraphics{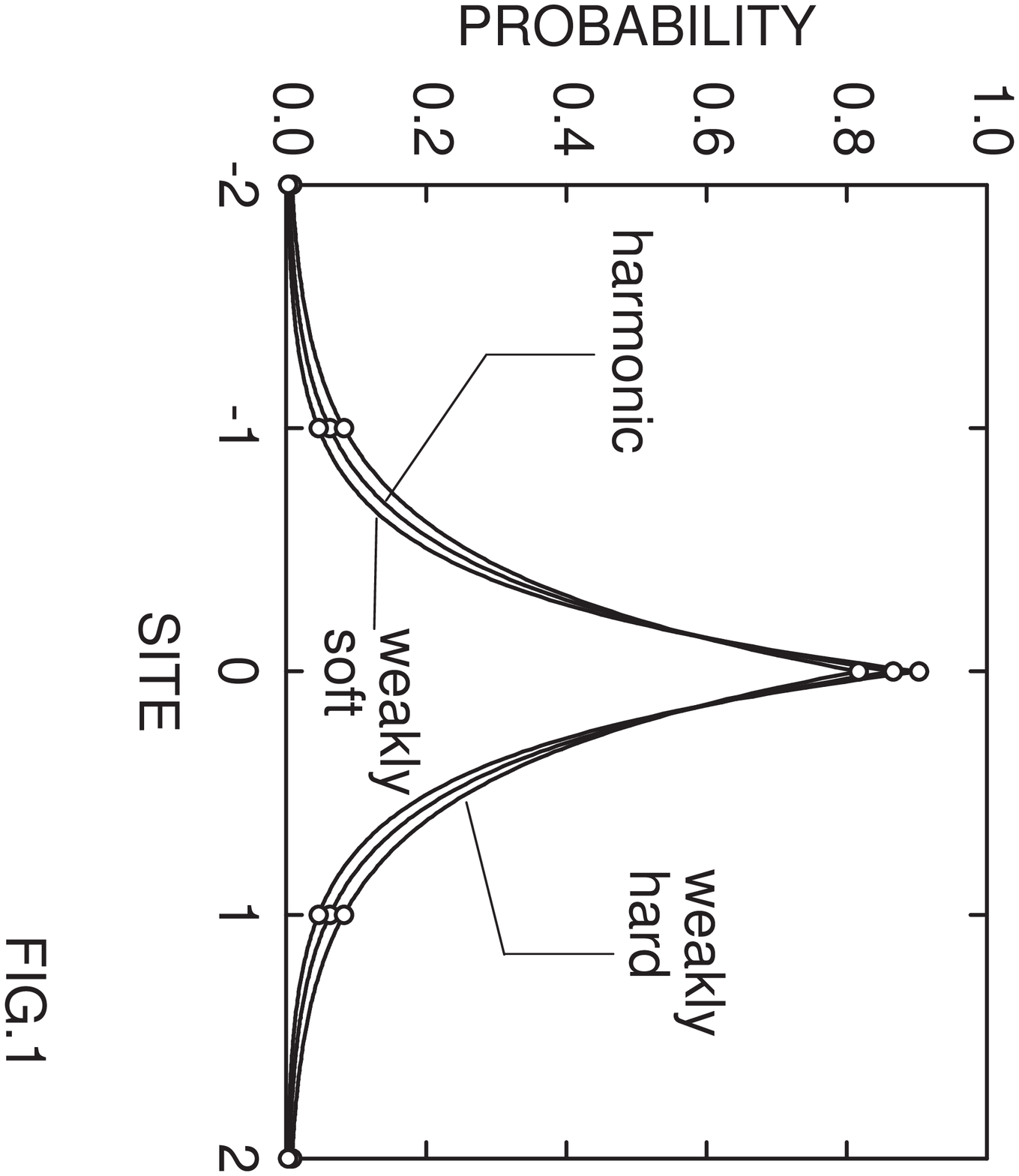}  }	
\end{center}
\end{figure}

\newpage
%
\begin{figure}[h]
\begin{center}
\leavevmode
\hbox{
\includegraphics{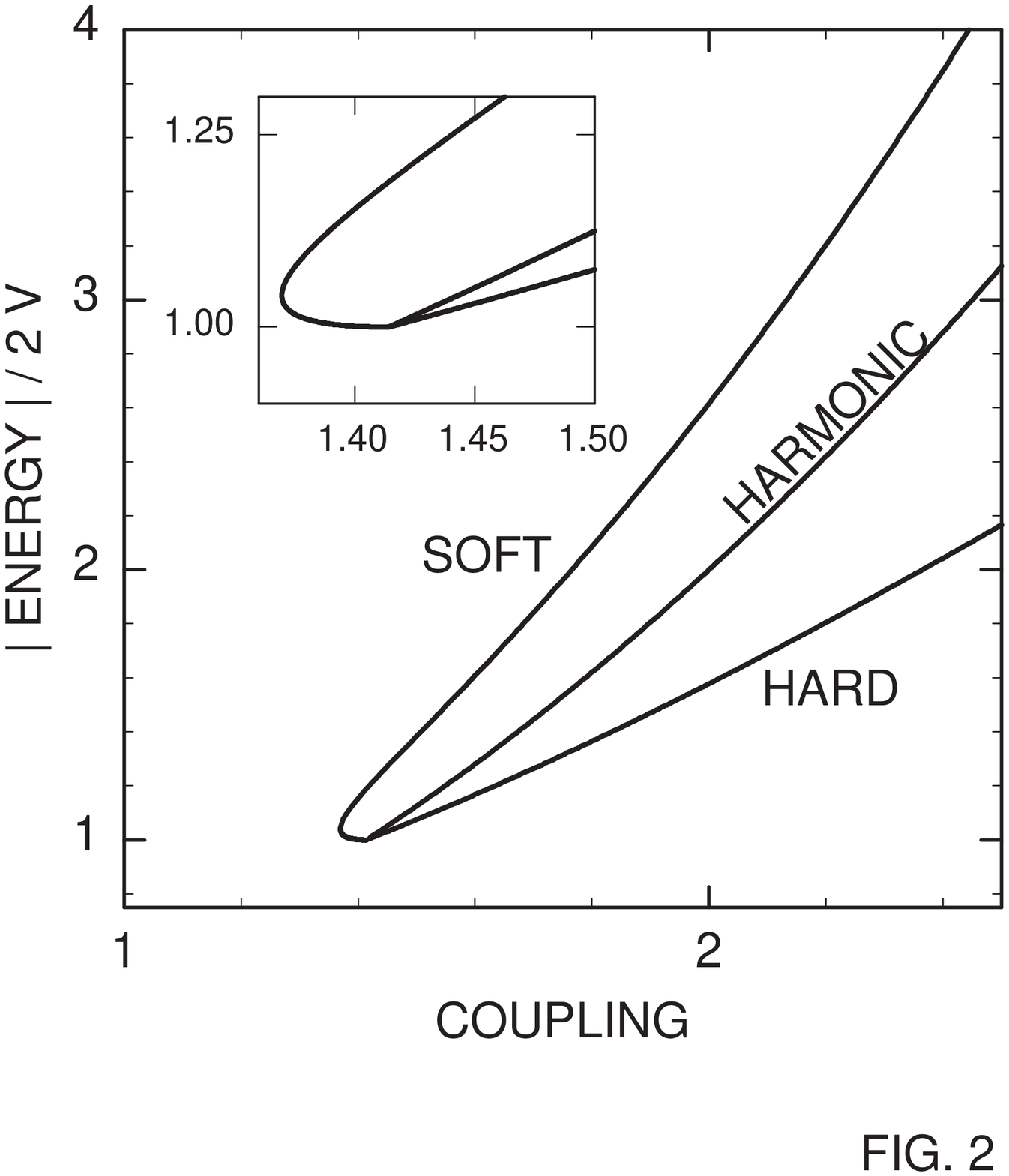}  }	
\end{center}
\end{figure}
\newpage
 
%
\begin{figure}[h]
\begin{center}
\leavevmode
\hbox{
\includegraphics{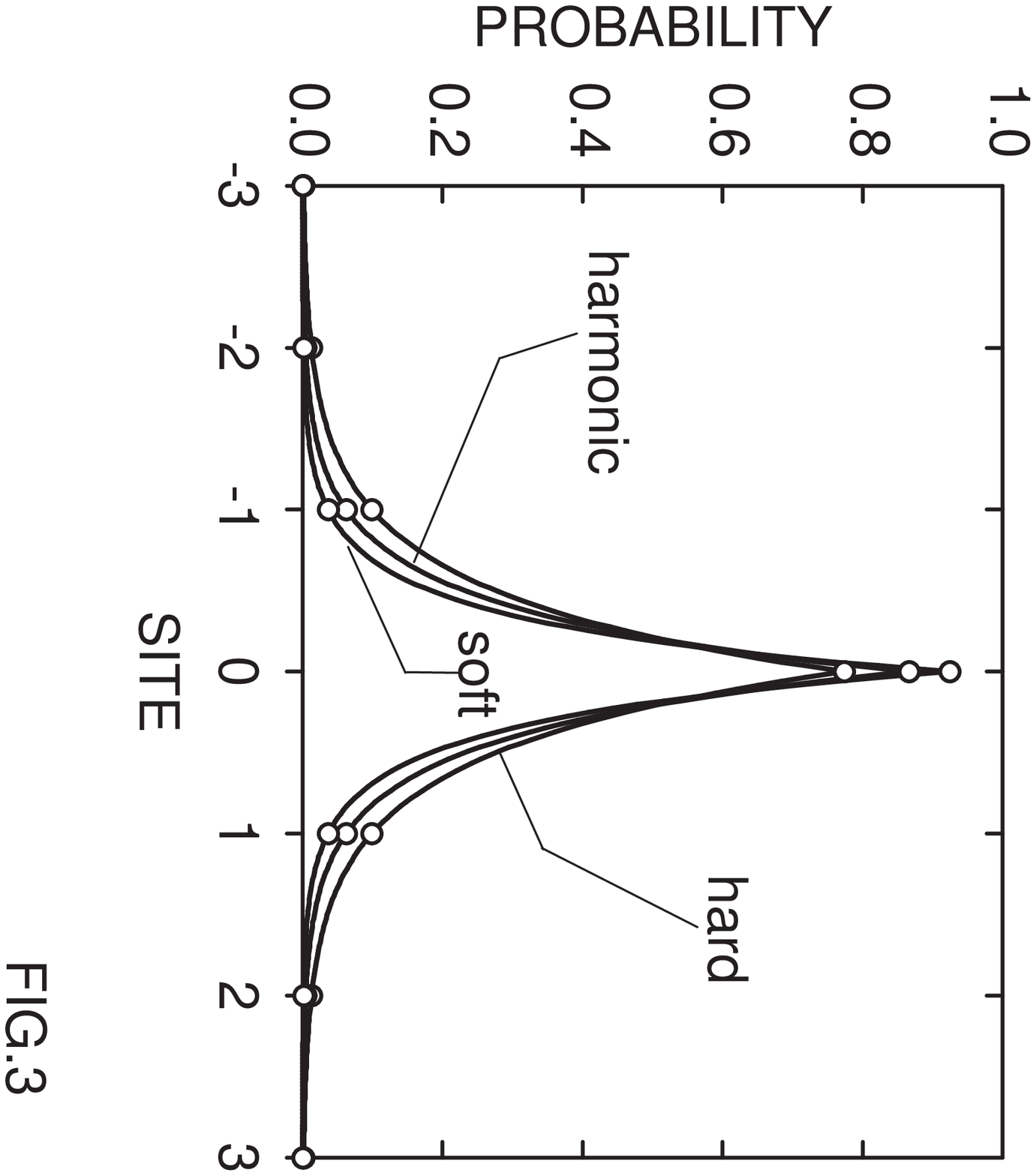}  }	
\end{center}
\end{figure}
\newpage

%
\begin{figure}[h]
\begin{center}
\leavevmode
\hbox{
\includegraphics{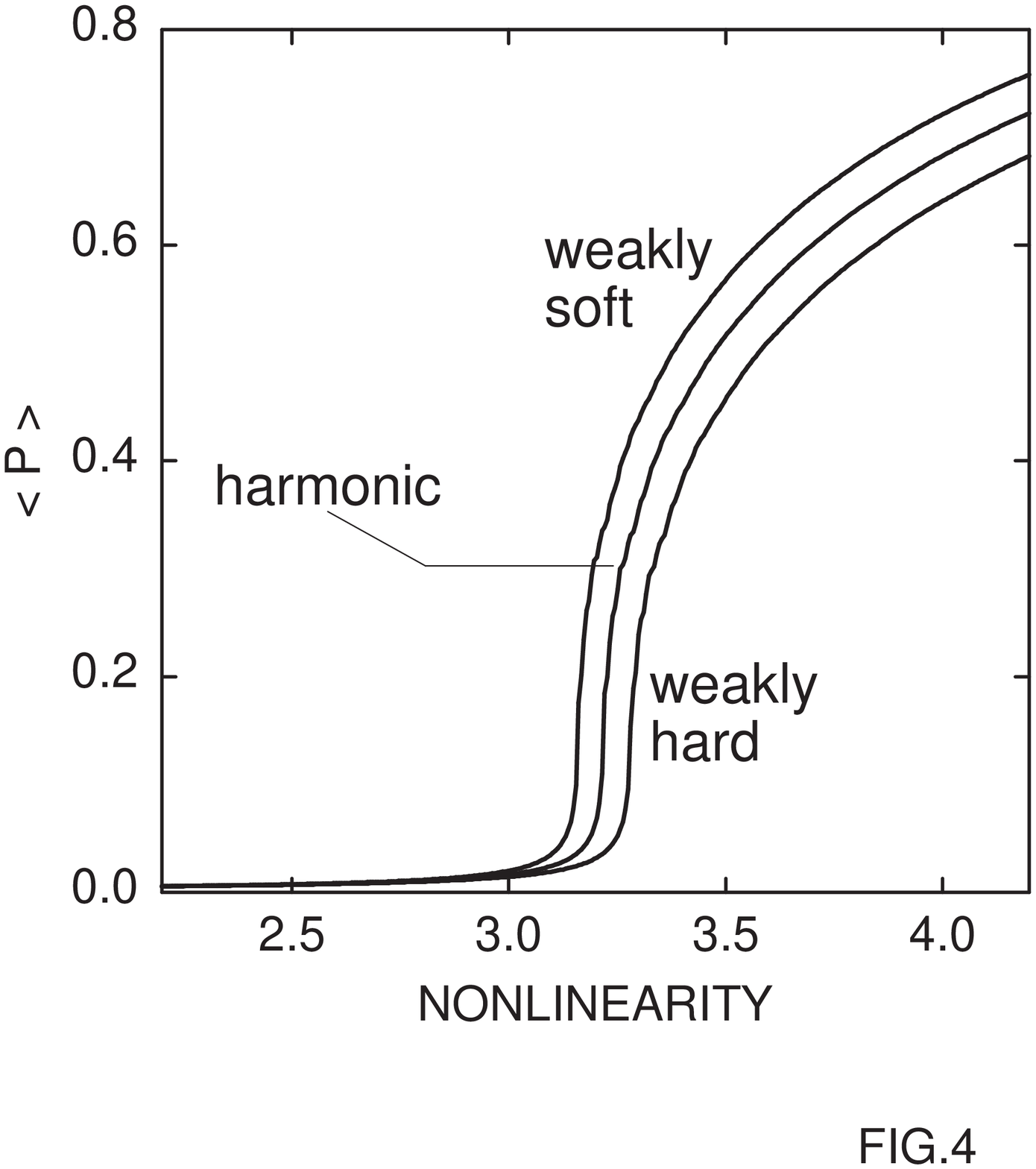}  }	
\end{center}
\end{figure}
\newpage

%
\begin{figure}[h]
\begin{center}
\leavevmode
\hbox{
\includegraphics{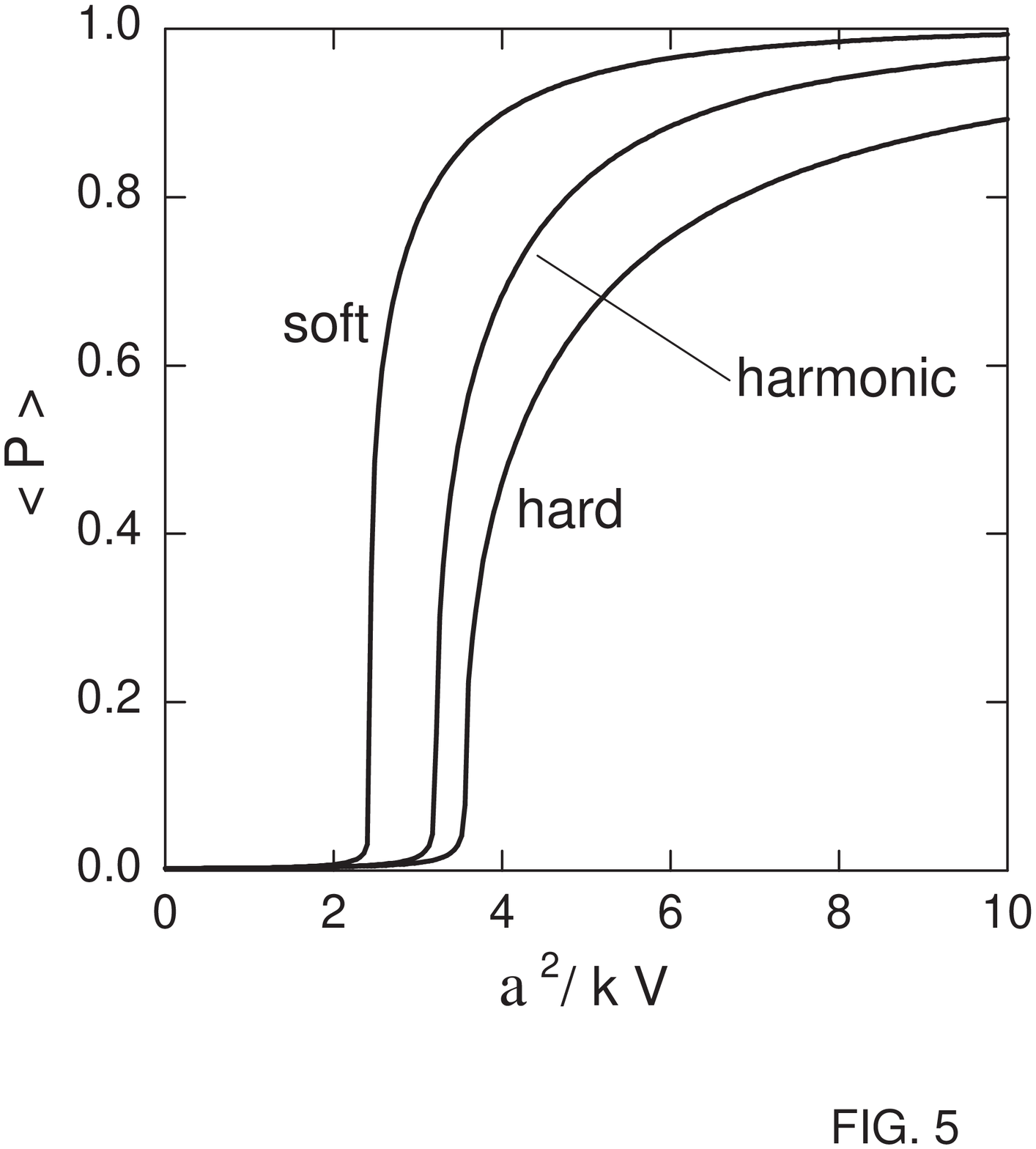}  }	
\end{center}
\end{figure}


\begin{thebibliography}{9}
\bibitem{comment}
These two features, lower critical nonlinearity and a more abrupt selftrapping
transition are of interest in nonlinear optics, for the problem of designing an
all-optical switch.
\end{thebibliography}
\end{document}